\documentclass{PoS}

\title{Observational Signatures of Lyman Alpha Emission from Early Galaxy Formation.}

\ShortTitle{Ly-$\alpha$ Emission from Early Galaxy Formation.}

\author{\speaker{Mark DIJKSTRA}\\%
        Max Planck Institute for Astrophysics\\
        E-mail: \email{dijkstra@mpa-garching.mpg.de}}

\abstract{The next generation of telescopes aim to directly observe the first generation of galaxies that initiated the reionization process in our Universe. The Ly$\alpha$ emission line is robustly predicted to be the most prominent intrinsic spectral feature of these galaxies, making it an ideal target to search for and study high-redshift galaxies. I briefly discuss why Ly$\alpha$ emitting galaxies (LAEs) provide a good probe of the intergalactic medium (IGM) during the Epoch of Reionization (EoR). I argue that if we wish to fully exploit LAEs as a probe of the EoR, it is important to understand what drives their observed redshift evolution after reionization is completed. One important uncertainty in interpreting existing LAE observations relates to the impact of the ionized IGM on Ly$\alpha$ photons emitted by galaxies, which is strongly connected to the effects of scattering through HI in galactic outflows.  Scattering through galactic outflows can also modify the Ly$\alpha$ spectral line shape such that $> 5\%$ of the emitted Ly$\alpha$ radiation is transmitted directly to the observer, even through a fully neutral IGM. Finally, I discuss what is required --observationally and theoretically-- to resolve the uncertainties that affect existing interpretations of data on LAEs.}

\FullConference{Cosmic Radiation Fields: Sources in the early Universe - CRF2010,\\
		November 9-12, 2010\\
		Desy Germany}
		
\begin{document}

\section{Introduction: Ly$\alpha$ Emitting Galaxies as a Probe of the Epoch of Reionization}
\begin{figure}
\begin{centering}
\includegraphics[width=.68\textwidth,angle=90]{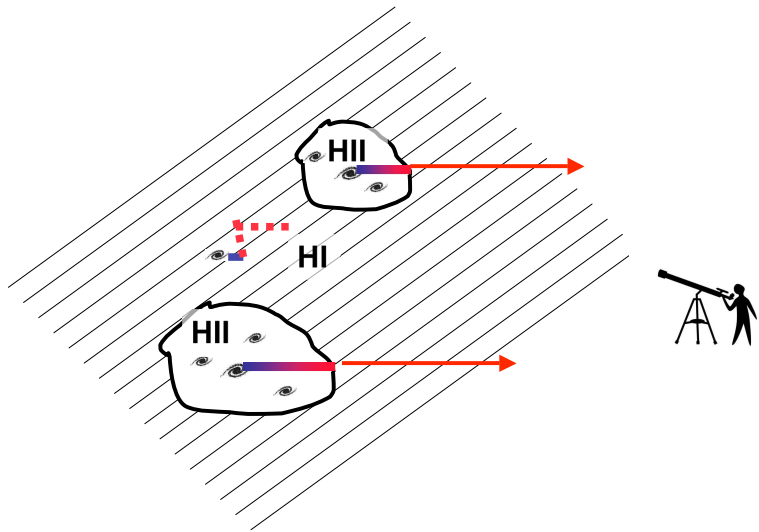}
\caption{This Figure shows schematically why Ly$\alpha$ emitting galaxies (LAEs) probe the distribution of neutral intergalactic gas during the EoR. Ly$\alpha$ photons emitted by galaxies inside large HII regions can redshift away far from the line resonance before they enter the neutral IGM (as indicated by the {\it color-changing solid lines}). As a result of this redshift, some of these photons can propagate freely to the observer. However, for galaxies inside smaller HII regions all Ly$\alpha$ photons scatter through the neutral IGM (represented by the {\it dotted lines}) into a very low surface brightness `fuzz' that is much too faint to be detected with existing telescopes \cite{LR99,DW10}.  Because the neutral IGM affects the detectability of Ly$\alpha$ photons, we expect the reionization process to leave an imprint in various statistics (number counts, clustering, ...) of LAEs \cite{HS99,McQ}.}
\end{centering}
\label{fig:schemelae}
\end{figure}

The Ly$\alpha$ emission line is robustly predicted to be the most prominent intrinsic spectral feature of the first generation of galaxies that initiated the reionization process in our Universe. The Ly$\alpha$ line can be heavily suppressed by intervening, neutral intergalactic gas. As a result, the process of reionization leaves an imprint on various statistics of Ly-$\alpha$ emitting galaxies (Fig~1, \cite{HS99}). However, {\it if we wish to fully exploit Ly$\alpha$ emitters (LAEs) as a probe into the Epoch of Reionization (EoR), it is important to understand what drives their observed redshift evolution after reionization is completed}. Otherwise, it is difficult to tell what other parameters are important in driving the redshift evolution of LAEs, and whether these parameters can be evolving during the EoR as well. I argue that one of the key uncertainties in interpreting existing LAE observations relates to the impact of the ionized intergalactic medium (IGM) on Ly$\alpha$ photons emitted by galaxies.

\section{The Opacity of the Ionized IGM to LAEs.}

\begin{figure}
\includegraphics[width=.7\textwidth,angle=90]{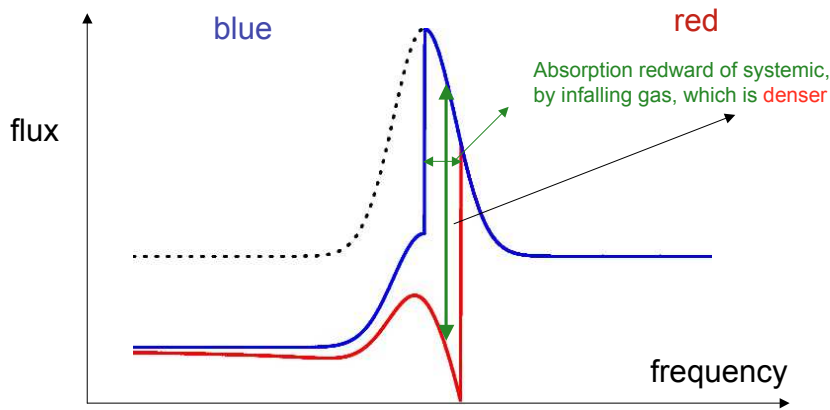}
\caption{This Figure shows schematically how the `local' IGM (or CGM) affects an initially Gaussian Ly$\alpha$ emission line (represented by the {\it dotted line}). To first order, the impact of the IGM may be captured by suppressing the blue half (i.e. shorter wavelengths) by a constant factor $\exp[-\tau_{\rm IGM}]$ which is derived from observations of the Ly$\alpha$ forest (represented by the {\it blue line}). However, the gravitational potential of the dark matter halo which hosts the LAE attracts the surrounding gas, which is denser than average \cite{Barkana}. Infall of this overdense nearby gas can suppress the observed Ly$\alpha$ line at frequencies significantly redward of the Ly$\alpha$ resonance \cite{preIGM,Santos}. As a result,  even the ionized IGM may transmit as little as $\sim 10-30\%$ of the Ly$\alpha$ photons \cite{IGM,D10,ZZ1,Laursen}. However, this number depends on how galactic winds affect the transport of Ly$\alpha$ photons through the ISM \cite{Santos,V08,DW10}.}
\label{fig:infall}
\end{figure}

The opacity of ionized gas in close (few tens to few hundreds of kpc) proximity to galaxies can be very opaque (see Figure~2, \cite{Santos,preIGM,IGM}). This has been confirmed repeatedly with numerical simulations \cite{D10,Laursen,ZZ1}. The opacity of this gas can fluctuate from sightline-to-sightline because of the inhomogenous distribution of the gas, and is subject to radiation fields that are dominated by nearby sources \cite{IGM}. Understanding the opacity of the `local' intergalactic medium (or circum galactic medium) is crucial to interpreting observations of LAEs before and after reionization has been completed. This is because from an observational point of view, the opacity of the local IGM is completely degenerate with the dust opacity in the interstellar medium: that is, it is irrelevant to existing observations whether Ly$\alpha$ photons are scattered into a low surface brightness fuzz that is too faint to have been detected, or whether they are absorbed by dust grains and then re-radiated in the infrared. The end result of both processes is effectively a suppressed Ly$\alpha$ flux from the source. Any uncertainty in the local IGM opacity thus translates to an equally large uncertainty in the dust opacity. To better understand the local IGM opacity, it is very important how galactic winds affect Ly$\alpha$ transport (see \S~3). 

\section{The Importance of Galactic Winds}
\begin{figure}
\begin{centering}
\includegraphics[width=.6\textwidth,angle=90]{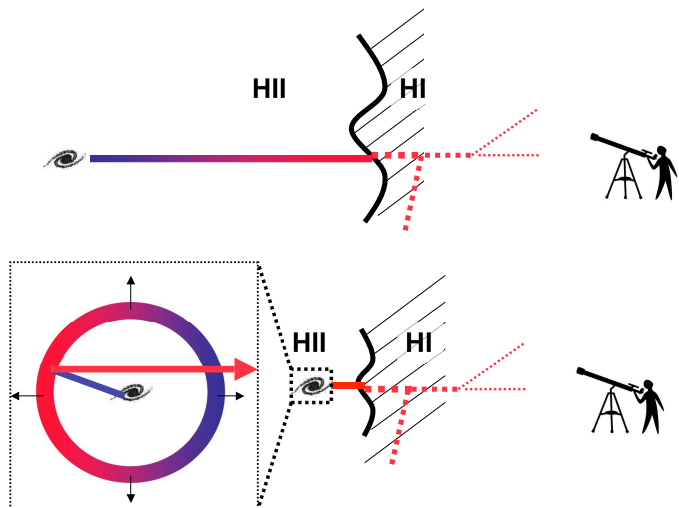}
\caption{Schematic explanation for why outflows promote the detectability of Ly$\alpha$ emission from galaxies surrounded by significant amounts of neutral intergalactic gas: The {\it top panel} is the same as Figure~1). The {\it bottom panel} shows that outflows surrounding star forming regions (represented by the expanding ring. The far side is receding form the observer and has a larger redshift, which is represented by the color) can Doppler boost Ly$\alpha$ photons to frequencies redward of the resonance. In this scenario, photons are `launched' far into the red wing of the line, where the IGM opacity is reduced. As a result, a fraction of Ly$\alpha$ photons can propagate directly to the observer even without a large H II bubble (taken from \cite{DW10}). Winds can also strongly reduce the opacity of the `local' IGM (or CGM)  because they can scatter photons to frequencies where even infalling gas cannot affect them \cite{DW10}.}
\end{centering}
\label{fig:scheme}
\end{figure}

\begin{figure}
\begin{centering}
\includegraphics[width=.33\textwidth,angle=270]{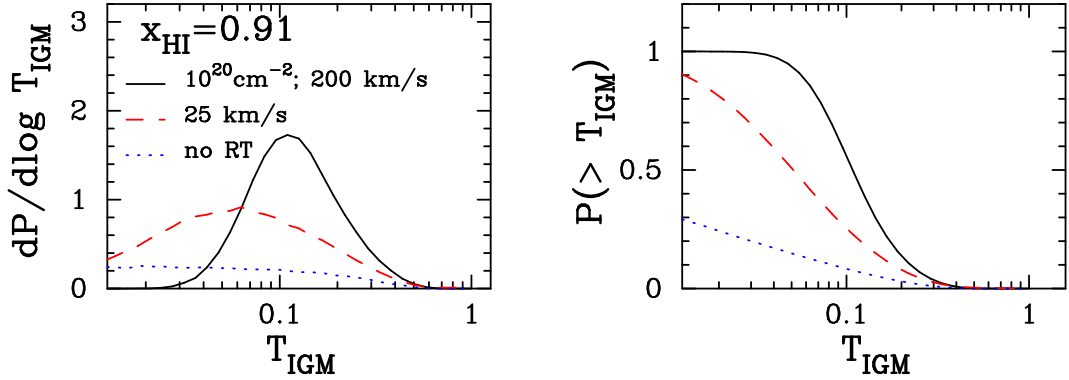}
\caption{We show the probability density function (PDF) ({\it left panel}) and the cumulative distribution function (CDF) ({\it right panel}) of $T_{\rm IGM}$-the fraction of photons that is transmitted through the IGM- for a Universe with a volume filling factor of neutral hydrogen of $x_{\rm HI}=0.91$.  Our calculations properly account for infall in the `local' IGM (or CGM), and for the large scale distribution of neutral intergalactic gas. We focuss on galaxies populating dark matter halos of $M_{\rm halo}\sim 10^{10} M_{\odot}$ at $z=8.6$.  The {\it red dashed line} ({\it black solid line}) shows the model with $v_{\rm wind}=25$ (200) km s$^{-1}$, and $N_{\rm HI}=10^{20}$ cm$^{-2}$ in both models. The {\it blue dotted line} shows a model with no winds (see text). This Figure illustrates ({\it i}) that the IGM can transmit a significant fraction of  Ly$\alpha$ photons, despite the fact that reionization has only just started, and ({\it ii}) that the IGM becomes even more `transparent' when Ly$\alpha$ scattering in the ISM is affected by galactic winds (from Ref~\cite{prep}).}
\label{fig:pdf}
\end{centering}
\end{figure}

Galactic winds can Doppler boost Ly$\alpha$ photons to frequencies redward of systemic velocity of a galaxy. This is illustrated pictorially in the {\it bottom panel} of Figure~3. In this picture, photons can be `launched' far into the red wing of the line by scattering off a galactic outflow \cite{V08}. At these frequencies, even infalling local intergalactic gas cannot affect them \cite{DW10}. This effect can also be important during the EoR: a wind-imparted redshift can cause a non-negligible fraction of Ly$\alpha$ photons to propagate directly to the observer even without a large H II bubble (taken from \cite{DW10}). We have quantified this is more detail in Figure~4 where we show the probability density function (PDF, {\it left panel}) and the cumulative distribution function (CDF, {\it right panel}) of $T_{\rm IGM}$-the fraction of photons that is transmitted through the IGM- for a Universe with a volume filling factor of neutral hydrogen of $x_{\rm HI}=0.91$. This Figure shows that the IGM can transmit a significant fraction of  Ly$\alpha$ photons even when reionization has only just started, especially when Ly$\alpha$ photons scatter off galactic winds (from Ref~\cite{prep}). Precisely how the IGM affects the visibility of Ly$\alpha$ photons, depends strongly on how important outflows are in the Ly$\alpha$ transfer process, before and after reionization has been completed.

\section{(Some) Observations and Models that are needed to move forward.}

While the presence of winds in galaxies is undisputed (e.g. \cite{Steidel10}), and while there are strong indication that winds play a role in the scattering process \cite{DW10}, 
the precise impact of the local (ionized) IGM is not well known. This is mostly because models have either focused entirely on scattering in the IGM \cite{IGM,ZZ1}, or entirely on scattering through galactic outflows \cite{V08,Laursen}. A next step on the theoretical side is to consider both processes simultaneously. Observationally, we foresee that the following observations are/would be very helpful in differentiating between scattering in the IGM and ISM (warning, this list is likely not complete):

\begin{itemize}
\item The redshift evolution of Ly$\alpha$ line shapes (as in e.g. ref \cite{Cas}). The IGM becomes more opaque with redshift, and we expect the IGM to leave some redshift dependent signature on Ly$\alpha$ line shapes.

\item The observed Ly$\alpha$ EW-PDF of drop-out galaxies as a function of $z$ and UV-magnitude \cite{Stark10}.

\item Measurements of the `volumetric Ly$\alpha$ escape fraction' as in ref \cite{Hayes10,Blanc}. 

\item Clustering measurements of larger samples of LAEs at z <6 to measure the strength of the so-called `Zheng effect' \cite{ZZ1}, the amplitude of which depends on galactic winds \cite{DW10}.

\item Ly$\alpha$ vs H$\alpha$ line shapes (as in ref \cite{Fink}, also see ref \cite{McL}). This constrains provides direct constraints on radiative transfer models, and the first such measurements have been reported.

\item Polarization: Ly$\alpha$ photons scattered by galactic outflows may obtain high levels of linear polarization, while this is less so for photons that are resonantly scattered in the IGM \cite{DL08}.
\end{itemize}

Many of these observations are feasible with existing instruments, which makes this an exciting research area. 

{ \bf Acknowledgements} I thank the organizers for organizing this workshop, and for the opportunity to present this work.

\end{document}